\documentclass[paper]{elsarticle}

\usepackage{lineno,hyperref}
\usepackage{graphicx}
\modulolinenumbers[250]

\journal{New Astronomy Reviews}

\begin{document}

\begin{frontmatter}

\title{What Regulates Galaxy Evolution?  \\Open Questions in our Understanding of Galaxy Formation and Evolution.}

\author[mymainaddress]{Gabriella De Lucia}
\ead{delucia@oats.inaf.it}

\author[mysecondaryaddress]{Adam Muzzin}
\ead{muzzin@strw.leidenuniv.nl}

\author[mysecondaryaddress]{Simone Weinmann}
\ead{weinmann@strw.leidenuniv.nl}

\address[mymainaddress]{INAF - Astronomical Observatory of Trieste, via
  G.B. Tiepolo 11, I-34143 Trieste, Italy}
\address[mysecondaryaddress]{Leiden Observatory, Leiden University, PO Box 9513
2300 RA Leiden, The Netherlands}

\begin{abstract}
In April 2013, a workshop entitled ``What Regulates Galaxy Evolution''
was held at the Lorentz Center. The aim of the workshop was to bring
together the observational and theoretical community working on galaxy
evolution, and to discuss in depth of the current problems in the
subject, as well as to review the most recent observational
constraints. A total of 42 astrophysicists attended the workshop. A
significant fraction of the time was devoted to identifying the most
interesting ``open questions'' in the field, and to discuss how
progress can be made. This review discusses the four
questions (one for each day of the workshop) that, in our opinion,
were the focus of the most intense debate. We present each question in
its context, and close with a discussion of what future directions
should be pursued in order to make progress on these problems.
\end{abstract}


\end{frontmatter}

\linenumbers

\section{Introduction}

In the last decade, a number of observational tests of the standard
cosmological paradigm have ushered in a new era of ``precision
cosmology''. Our current standard model for structure formation is able
to reproduce simultaneously a number of important observational
constraints, ranging from the temperature fluctuations in the cosmic
microwave background, the power spectrum of low redshift galaxies, to
the acceleration of the cosmic expansion inferred from supernovae
explosions.  While the cosmological paradigm appears to be firmly
established, a theory of galaxy formation continues to be elusive, and
our understanding of the physical processes that determine the
observed variety of galaxies is at best rudimentary.  Although much
progress has been made, both on the theoretical and observational
side, understanding how galaxies form and evolve remains one of the
most outstanding questions of modern astrophysics.  In addition to
being an interesting question on its own right, galaxy formation also
has important implications for cosmological studies. Indeed, at least
some cosmological probes use galaxies as tracers (e.g. those based on
measurements of galaxy clustering).  A better understanding of the
galaxy formation process is therefore crucial in order to improve our
knowledge of the mass-energy content of the Universe.

These are exciting times to study galaxy formation: a wealth of
new data are expected from ongoing and planned photometric and
spectroscopic surveys of the local and more distant Universe, at
different wavelengths. In parallel, the field of computational
astrophysics has progressed rapidly thanks to increasing computational
power and to the development of more sophisticated numerical algorithms.

In April 2013, the authors of this paper organized a workshop at the
Lorentz Center
\footnote{http://www.lorentzcenter.nl/lc/web/2013/528/info.php3?wsid=528}
to bring together the observational and theoretical community and
discuss in depth of the current problems in galaxy formation, as well
as to review the most recent observational constraints. A total of 42
astronomers participated in the workshop, including theorists and
observers working on a wide range of topics in galaxy formation, from
dwarf galaxies, to massive galaxies, isolated galaxies and cluster
galaxies, from very high to very low redshift. A significant portion
of the workshop was devoted to identifying the most interesting open
questions in galaxy evolution, and how progress can be made on these
problems. Many interesting questions were debated. Below, we
provide a summary of the four questions (one for each day of the
workshop) that, in our opinion, were the focus of the most intense
debate. We will discuss those four questions in their context, and
close with an outlook on what areas in galaxy formation we believe are
especially promising to help making progress on the identified
problems. In particular, the four questions selected are: (i) Are we
  reaching a fundamental limit in our ability to measure properties such as
  stellar mass and star formation rates? (ii) What is the star formation and
  assembly history of galaxies with mass below $10^9\,{\rm M}_{\odot}$? (iii)
  Does the central-satellite division provide the right framework to study
  galaxy evolution? (iv) We understand which processes affect galaxies in
  different environments. Do the details matter?

Since they were selected simply based on the interest they generated,
the four questions are quite different in nature: (i) and (iii) are
technical, (iv) a somewhat philosophical one, and (ii) is more a
standard science question.

\section{Question 1 - Are we reaching a fundamental limit in our ability to 
  measure properties such as stellar mass and star formation rates?}

Two of the most fundamental parameters that describe a galaxy are its
total mass in stars, and the rate at which stellar mass grows via star
formation, the star formation rate (SFR).  Measuring the evolution of
stellar masses and SFRs both for individual galaxies and for the
Universe as a whole occupies a substantial fraction of the
observational resources devoted to the study of galaxy formation
\citep[see for example][just to mention a
  few]{Kauffmann2003,Brinchmann2004,Salim2007,vandokkum2010,Baldry2012,Muzzin2013b}. Given
their important role for assessing the success of theoretical models
\citep[e.g.][]{Delucia2007,Schaye2010,Weinmann_etal_2012}, increasing
the precision with which stellar masses and SFRs are measured, over a
wide range of redshifts and halo masses, continues to be a major goal
of the observational community.

Over the last two decades, incredible progress has been made in
obtaining high-quality data for this purpose.  In particular, the
Sloan Digital Sky Survey (SDSS) has provided high-quality photometry
and spectroscopy which have allowed the measurement of stellar masses
and SFRs for millions of galaxies
\citep{Kauffmann2003,Brinchmann2004,Blanton2005}.  While no survey
complementary to the SDSS  exists for the high-redshift Universe yet,
the coming of wide-field NIR and MIR cameras, the WFC3 camera on HST,
as well as significant improvements in photometric redshift techniques
have opened up studies of the stellar masses of samples of up to
hundreds of thousands of galaxies, up to as far as z $\sim$ 8
\citep[e.g.,][]{Marchesini2009,Labbe2010,Muzzin2013b,Ilbert2013}.
Likewise, access to the FIR and Sub-mm from $Spitzer$, $Herschel$, and
now $ALMA$ have allowed us to study dusty star formation up to $z
\sim$ 6 \citep[e.g.,][]{Chapman2005,Daddi2007,Riechers2013}, and
$GALEX$ has made the study of SFRs from the rest-frame UV available in
the local Universe
\citep[e.g.,][]{Martin2005,Schiminovich2005,Salim2007}.

With this extraordinary increase in the sample sizes and data quality
for distant galaxies, it has becoming increasingly clear that the
dominant source of uncertainty is provided by {\it systematics} in the
conversion of the photons we observe, into physical quantities
\citep{Conroy2013}. Without an improvement in our understanding of
these systematic uncertainties, it is unclear whether we will be able
to take advantage of the nearly overwhelming samples of galaxies that
will be available for study from surveys with upcoming telescopes such
as {\it LSST}, {\it Euclid}, and {\it WFIRST}. Can we really develop techniques to
reduce systematic uncertainties in deriving key quantities such as
stellar mass and SFRs, or are we truly reaching a fundamental limit in
our ability to do so?  

Stellar masses are typically determined for galaxies by fitting their
spectral energy distributions (SEDs) measured from either spectra, or
broadband photometry to synthetic spectra derived from stellar
population synthesis (SPS) codes.  A thorough discussion of this
process, and the inherent challenges with it can be found in the
recent review by \cite{Conroy2013}.  In brief, SPS models encode the
current state-of-the-art knowledge of stellar evolution both on and
off the main-sequence, and use isochrones combined with both real and
synthetic spectra for stars to produce composite SEDs that the data
can be fit to.  Not all SPS models are alike, with each employing
slightly different isochrones and/or treatment of the various phases
of stellar evolution.  Therefore, for identical raw observational
data, different stellar masses are derived using different SPS codes.
The difference between SPS codes was recently highlighted by the
various treatments of the thermally-pulsating asymptotic branch phase
(TP-AGB) of stellar evolution (see Section \ref{sec:futurespmodels}).
This is challenging to model, yet can have a large effect on derived
synthetic SEDs.  Because of the TP-AGB phase, and other differences
between the codes, most recent observational studies have concluded
that the largest systematic uncertainty in deriving stellar masses
currently is the uncertainty in how to treat stellar evolution (i.e.,
the SPS codes themselves).

\begin{figure}
\begin{center}
\includegraphics[width=1.0\textwidth]{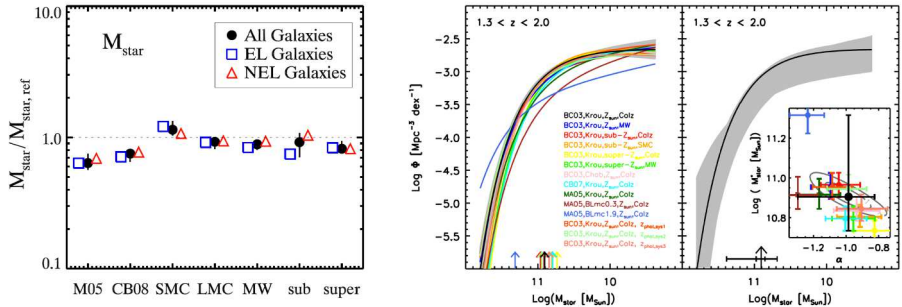}
\caption{Left panel: The effect of varying assumptions in SED fitting
  on the derived stellar mass of individual galaxies at $z \sim$ 2
  compared to the default model (see text) from \cite{Muzzin2009}.
  All galaxies have 13 band photometry and spectroscopic redshifts.
  EL and NEL denote galaxies with and without emission lines
  (effectively quiescent or star-forming galaxies).  The largest
  systematic uncertainties come from the choice of SPS model.  Right
  panel: The effect of varying assumptions in SED fitting on the
  stellar mass function of galaxies at $z \sim$ 1.7 from
  \cite{Marchesini2009}.  Other than unusual IMFs, again the largest
  systematic uncertainty is from the choice of SPS model.}
\label{fig:sysmass}
\end{center}
\end{figure}

This is illustrated in the left panel of Figure~\ref{fig:sysmass}
(from \cite{Muzzin2009}) that shows the effect of varying assumptions
parameters in the SED fitting to determine the systematic differences
in the derived stellar mass of individual massive galaxies at $z \sim
2$.  Parameters were varied relative to a default template set:
Bruzual \& Charlot (2003) SPS models, the Calzetti dust law, and solar
metallicity.  Figure~\ref{fig:sysmass} shows that the largest
systematic uncertainty in the determination of stellar masses for
individual galaxies is the choice of SPS model, and this difference is
a factor of $\sim$ 1.6.  The right panel of Figure~\ref{fig:sysmass}
(from \cite{Marchesini2009}) shows the same approach but this time the
effect on the full stellar mass function at $z \sim$ 1.7.  The most
extreme effect on the stellar mass function is the use of bottom light
initial mass functions (IMFs). Thereafter, the next largest
  effect is the choice of the SPS model. Note that a bottom light IMF
  is disfavoured by more recent data \cite[e.g.][and references therein]{Conroy_vanDokkum2012,Shetty_Cappellari2014}.

Our understanding of stellar evolution is not the only limitation in
deriving stellar masses.  When using SPS models, additional free
parameters determine the output synthetic spectra such as the
metallicity of the stellar population, the IMF, the star formation
history (SFH), the dust attenuation law, and the dust geometry.  In
the case of fitting broad band photometry, which is not capable of
constraining these properties, usually a single metallicity (typical
solar), a single IMF (frequently Kroupa or Chabrier), a single dust
law (frequently Calzetti or the Milky Way), and a single dust geometry
(homogeneous screen) are assumed.  The SFH is now one parameter that
is commonly fit for, although most frequently it is assumed to have
simple functional forms such as declining exponentials (SFR $\propto$
$e^{-t/\tau}$), the so-called ``$\tau$-models".  More recently it has
been pointed out that $increasing$ $\tau$-models (i.e., SFR $\propto$
$e^{t/\tau}$) may be more appropriate for star-forming galaxies
\citep{Maraston2010,Papovich2011}, and this has been adopted for some
stellar mass determinations \citep{Lee2012}.  It is clear that these
myriad assumptions will underlie significant systematic uncertainties
in the stellar mass determinations as it is well-known that galaxies
have a range of metallicities, IMFs, SFHs, and dust geometries.

These issues are well known and have been for some time; however, it
is not clear how large the systematic uncertainties truly are because
there is no sample of galaxies for which we accurately know the
stellar mass from an independent method (e.g., star
counts). Therefore, other than globular clusters, there are few
populations that can be used as benchmarks for methods of stellar mass
determination.  Studies have tried varying as many of the parameters
in SED fitting as possible in order to make estimates of the current
level of systematic uncertainties using both real 
\citep[e.g.,][]{Maraston2006,Wuyts2007,Muzzin2009,Marchesini2009} and
simulated galaxies
\citep[e.g.,][]{Wuyts2009,Pforr2013,Mitchell2013}. However,
these are simply varying the so-called ``known unknowns'' and so
realistically provide only a lower-limit on the size of systematic
uncertainties given that there are certainly still ``unknown unknowns''
in the process (e.g. if unusual IMFs or strange dust laws exist, the
effect of these on deriving stellar masses has not been tested).

At the Lorentz workshop as part of one of the discussion sections,
M. Franx conducted a ``poll'' of the attendees asking what they felt
the current level of systematic uncertainties were in deriving stellar
masses.  Put another way, they were asked what they felt a ``safe''
error bar (random + systematic) would be for the stellar mass of a
typical galaxy.  While there were a range of opinions, the majority
felt that uncertainties were $\sim$ 0.5 dex for galaxies at $z \sim$
0, using SDSS spectra, and $\sim$ 1.0 dex for galaxies at $z \sim$ 2,
using standard broadband techniques.  This is substantially larger
than the systematic uncertainties derived from varying the ``known
unknowns'' above, but astronomers are cautious and well aware of
potential unknown unknowns.

A notable dissenter was C. Conroy, who suggested that uncertainties
were smaller, probably more like 0.3 dex locally, and $\sim$ 0.5 dex
at higher redshift.  He pointed out that evidence for this is the fact
that there exist scaling relations between the stellar mass of
galaxies and other properties such as their size, dynamical mass, and
SFR.  These parameters are determined in independent ways, and so if
the systematic uncertainties in stellar masses truly were 1.0 dex, it
is unlikely (although not impossible) that tight scaling relations
would exist.

To return more directly to the question at hand, which is, are we
reaching a fundamental limit in our ability to measure stellar masses?
The answer to the question has to be twofold, depending on how we
phrase exactly the question above. If the question is: are we reaching
a limit in our ability to improve our understanding of galaxy
formation using stellar masses determined from standard techniques?
Then, the answer must be ``yes''.  Other than a few unprobed areas of
parameter space for the stellar mass function, such as very high
redshift ($z >$ 4), or very low masses (Log(M$_{star}$/M$_{\odot}$)
$<$ 9.5 at $z >$ 1), the statistics and quality of photometric data
compared to that already in hand are unlikely to improve substantially
in the future.  More photometry cannot solve our problems.

In the most direct form of the question, which is are we truly reaching a
  ``fundamental limit'' (due to physical reasons) in our ability to measure
stellar masses, the answer has to be emphatically ``no''.  We should hope so
too, because if the uncertainties truly are 0.5 -- 1.0 dex as the attendees
suggested, then as a community we would like to think we could improve on such
a dismal situation!  We will come back to this in Section 6 below.

\section{Question 2 - What is the star formation and assembly history of 
  galaxies with mass below $10^9\,{\rm M}_{\odot}$?}

Galaxies with stellar mass below $10^9\,{\rm M}_{\odot}$ exhibit a variety of
physical properties, from dwarf spheroidals (dSph) that are gas poor and tend
to be concentrated around more massive galaxies, to dwarf irregulars (dIrr)
whose gas fraction can vary from zero to one, and tend to be more isolated than
dSphs. The star formation histories of both types of galaxies are very
stochastic, but they all contain a certain fraction of old stars
\citep{Weisz_etal_2011}. A relatively tight mass-metallicity relation is in
place which, together with information from gas masses, provide strong evidence
for significant metal and mass losses. Outflows from low-mass galaxies are
  likely metal-enriched (i.e. the outflows contain more metals per unit mass
  than the average interstellar medium)
\citep{Tremonti_etal_2004,Dalcanton_2007,Bouche_etal_2007}. Galaxies in this
mass regime provide very strong constraints to galaxy formation models, both in
terms of their total number density, and with respect to their physical
properties.

In fact, from the theoretical point of view, recent studies have
pointed out the existence of a {\it fundamental problem} with the
evolution of low mass galaxies in hierarchical galaxy formation
models, as well as in the state-of-the-art hydrodynamical simulations
\citep[see][and references therein]{Weinmann_etal_2012}. This problem
manifests itself in different forms: a dramatic difference is found
between the observed number density of low mass galaxies and that
predicted by galaxy formation models. When including a strong feedback
from supernovae, models are able to reproduce the observed galaxy
stellar mass function in the local Universe, but they consistently
over-produce the number density of sub-M$_*$ galaxies at higher
redshift \citep{Fontanot_etal_2009,Marchesini2009,Guo_etal_2011}. These same studies
also indicate that low-mass galaxies tend to be too old and passive
compared with observational measurements. Models also fail to
reproduce the observed anti-correlation between specific star
formation rates and stellar mass
\citep{Somerville_etal_2008,Firmani_etal_2010}. Finally, models
typically under-predict the observed specific star formation rates at
$z<2$, while over-predicting the same quantities at $z>3$
\citep{Daddi_etal_2007,Weinmann_etal_2011}.

Early attempts to address these problems focused on the simplified treatment of
satellite galaxies as the main problem responsible for (at least some of) the
disagreements mentioned. It is usually assumed that when a galaxy is accreted
onto a larger structure (i.e. when it becomes a satellite galaxy) its reservoir
of hot gas is stripped and therefore cannot replenish the galaxy with new fuel
for star formation \citep[this is the `strangulation' process discussed
  in][]{Larson_etal_1980}. Until a few years back, the common assumption was
that the stripping was happening instantaneously. This induced a very rapid
decline of the star formation histories of satellite galaxies, creating an
excess of red and passive galaxies with respect to the observations
\citep[e.g.][]{Weinmann2006b,Wang_etal_2007}. The exhaustion of star formation
occurs over a very short time-scale also because these models usually include a
very efficient stellar feedback, that removes any residual gas from the galaxy
adding it either to the hot component associated with the corresponding central
galaxy, or ejecting it outside the halo. In more recent studies, a more gradual
stripping of the hot gas reservoir has been assumed
\citep{Kang_vandenBosch_2008,Font_etal_2008,Weinmann_etal_2010,Guo_etal_2011}. Albeit
improved, the agreement with observational measurements is far from
satisfactory.

\citep{Fontanot_etal_2009} showed that the over-production of galaxies
at intermediate and low mass in the models is not solely due to an
incorrect treatment of the evolution of satellite galaxies. Rather,
this is mainly driven by an over-efficient formation of central
galaxies at high redshift, in haloes with circular velocities $\sim
100-200\,{\rm km}\,{\rm s}^{-1}$.  Therefore, mechanisms that only
affect satellite galaxies (such as strangulation or ram-pressure
stripping) or mechanisms that only affect low-mass haloes (such as
photoionizations) do not provide viable solutions. Suppressing the
formation of galaxies in small but compact haloes at high redshift is
not trivial: the density of the haloes is too high and their potential
wells are too deep to suppress star formation with heating from an
external UV background. Galactic winds certainly play an important
role, but they should not destroy galaxies of the same circular
velocity at lower redshift. As pointed out by H. Mo at the workshop,
it should also be noted that one cannot `hide' the mass ejected from
low-mass galaxies, because also the slope of the HI mass function is
very shallow. Indeed, attempts to solve the overabundance of low-mass
galaxies by reducing the star formation efficiency lead to a dramatic
over-prediction of the cold gas content of galaxies
\citep{Wang_etal_2012}.

At the time of writing, the consensus is that the discrepancies discussed above
require a critical revision of the feedback (and recycling) schemes that are
currently implemented in hierarchical galaxy formation models. In particular,
what appears to be needed is a mechanism that is able to decouple the growth of
low-mass galaxies (that occurs late) from that of their hosts (that, in
contrast, occurs early). As pointed out by \citep{Weinmann_etal_2012}, the
feedback schemes currently adopted in galaxy formation models are unable to
achieve this because of their dependence on halo mass and cosmic time. The same
authors stressed that a potentially important mechanism is that of gas
recycling. In current models, this is parametrized in a way that its efficiency
increases at higher redshift so that the net outflow rates are low at early
cosmic times and increase at lower redshift. Recent hydrodynamical simulations
suggest the evolution with cosmic time should be weaker
\citep{Oppenheimer_etal_2010}. In a recent study, \citep{Henriques_etal_2013}
modified the recycling scheme making gas re-incorporation time scales dependent
on halo mass and independent on redshift. This modification appears to move
model predictions closer to the observational data for the galaxy stellar mass
function at different redshift. The predicted passive fraction of low mass
dwarfs remains, however, too high with respect to observational
measurements.  Alternative scenarios that invoke suppression of gas
  accretion in low mass haloes can alleviate significantly the problem
  \citep{Bouche_etal2010,Lu_etal2014}. Unfortunately, however, these scenarios
  appears to be `ad hoc', and the exact mechanism remains unclear.

It is interesting to note that the excess of low and intermediate mass galaxies
is connected to a long-standing problem in the framework of hydrodynamical
simulations: gas cooling is very efficient at high redshift and in small
  and compact haloes. Thus, baryons condense early in clumps that then fall
  into larger haloes and merge via dynamical friction. This produces a net and
  significant transfer of angular momentum from the baryons to the dark
  matter, resulting in spiral galaxies with large bulges and compact disks
(this is the so called `angular momentum catastrophe'). The formation of a
realistic rotationally supported disk galaxy in a fully cosmological simulation
is still an open problem. Recent numerical work shows that it is in part due to
limited resolution, and related numerical effects that cause artificial angular
momentum loss and spurious bulge formation (for a detailed discussion, see
\citep{Mayer_etal_2008}). Feedback driven by supernovae explosions represents a
crucial ingredient to regulate the assembly of galaxies and avoid catastrophic
losses of angular momentum. Despite much progress, however, the formation of
thin and high angular momentum stellar disks remains a challenge \citep[][and
  references therein]{Scannapieco_etal_2012}. As in semi-analytic studies, some
form of `early feedback' seems to be required in order to suppress excessive
formation of stars at high redshift \citep{Stinson_etal_2013}.

But how well do we know the star formation and assembly histories of
these low-mass galaxies?

\begin{figure}
\begin{center}
\includegraphics[width=0.45\textwidth]{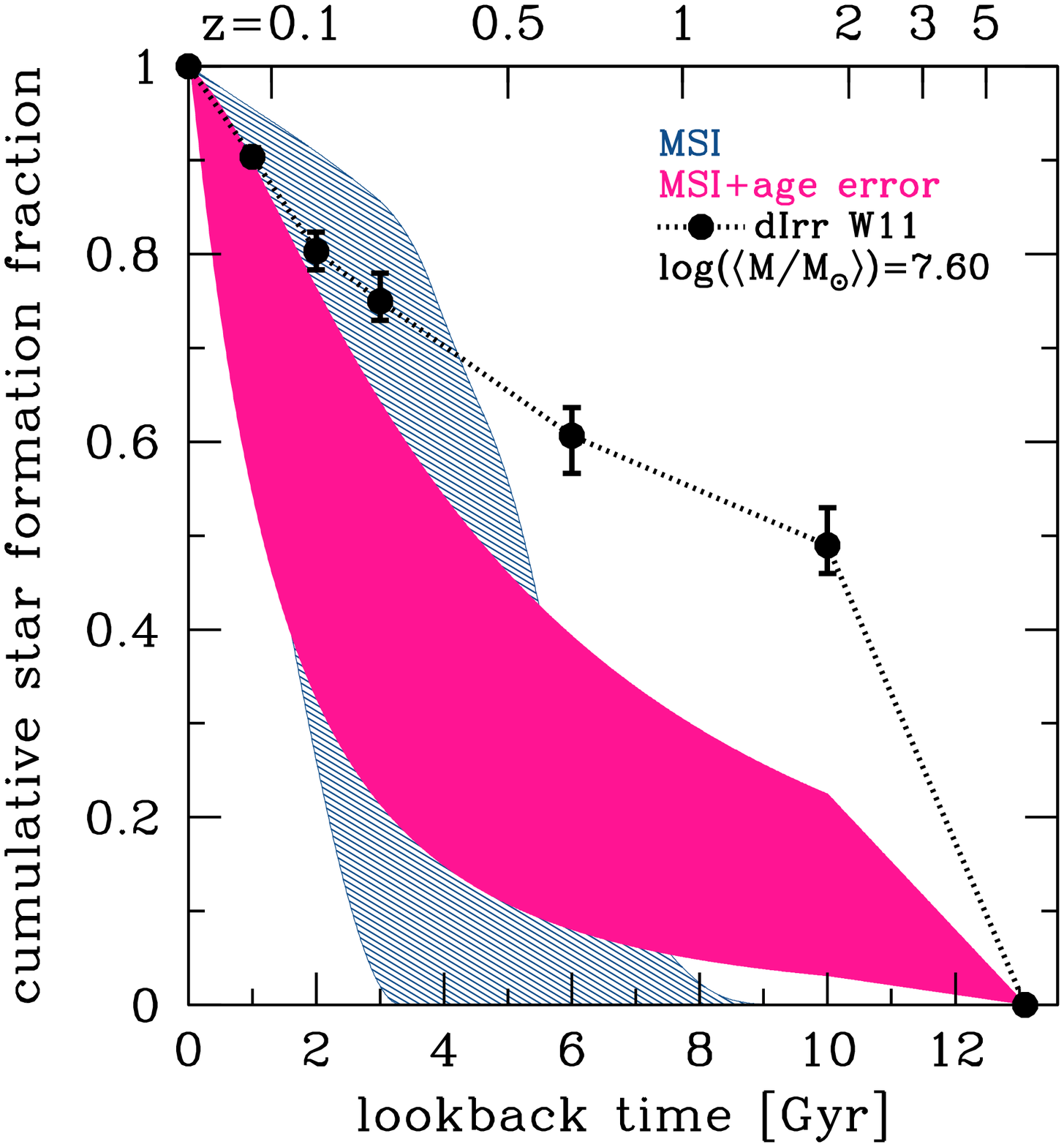}
\includegraphics[width=0.45\textwidth]{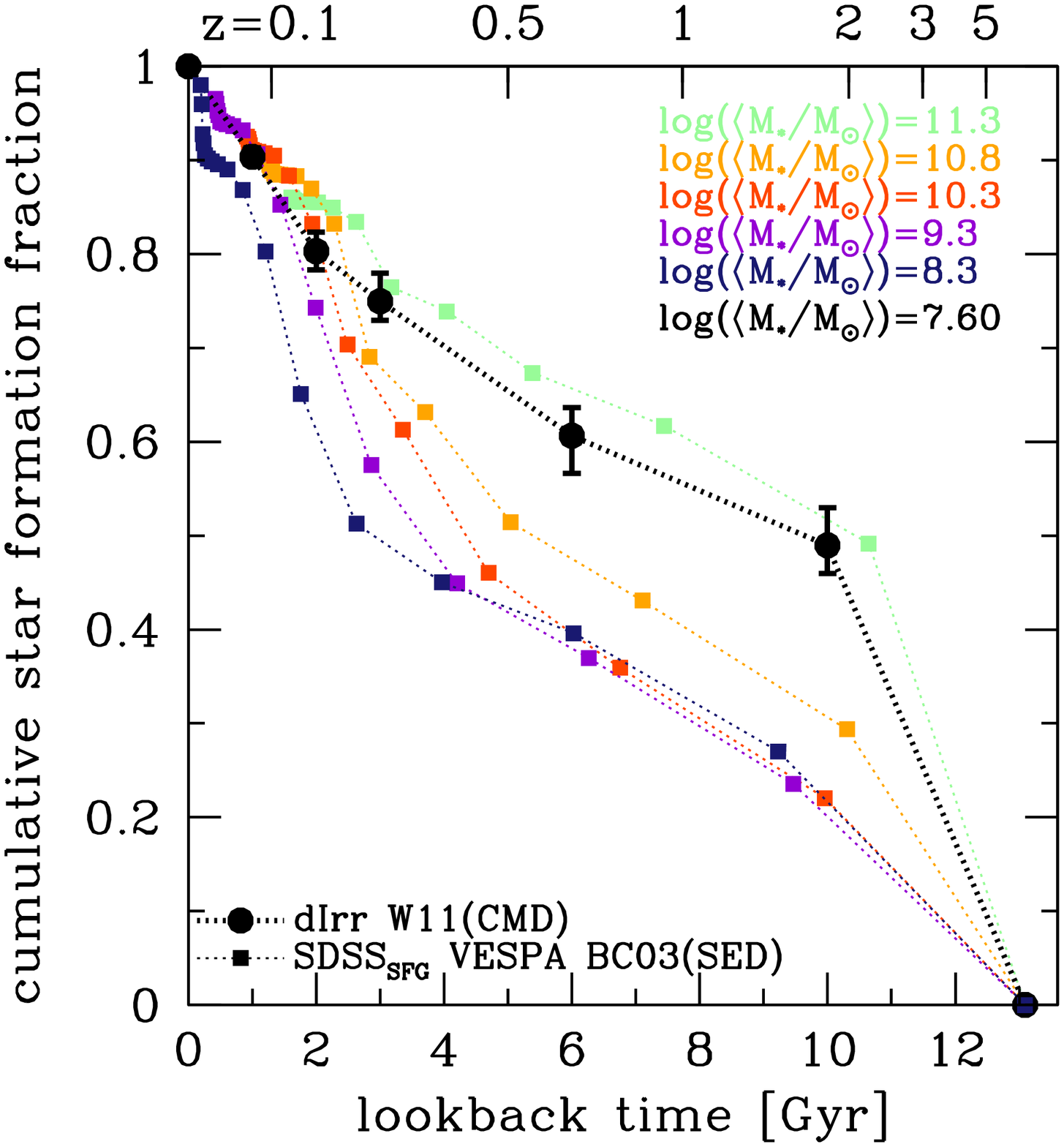}
\caption{Figures from \citep{Leitner_2012}. Left panel: Cumulative
  star formation from the MSI method (shaded regions). These are
  consistent with SED-fitting estimates when uncertainties on ages are
  accounted for. MSI trends are extrapolated over this mass
  range. Right panel: Cumulative star formation based on SED fitting
  for SDSS star forming galaxies in different mass bins. In both
  panels, the black points are CMD-based results for dwarf irregulars
  from \citep{Weisz_etal_2011}.}
\label{fig:leitner}
\end{center}
\end{figure}

As discussed in the previous section, a galaxy star formation history
can be constrained comparing its spectral energy distribution (SED) to
a model spectra obtained convolving different simple stellar
populations with a library of SFHs. This requires a number of
assumptions on e.g. the IMF, stellar evolution, dust
extinction, and chemical evolution. All these ingredients contribute
to making the uncertainty on SED derived star formation histories very
large (with the largest contribution coming from the assumed stellar
evolutionary tracks).  For local galaxies, the star formation history
can be independently constrained from the colour-magnitude diagram
(CMD). A figure shown at the workshop by J. Dalcanton received much
attention, and was the subject of much discussion:
Figure~\ref{fig:leitner} compares estimates of star formation
histories obtained from a SED fitting method, and the CMD diagram. In
particular, the left panel shows the cumulative star formation
fraction obtained using an approach (the `Main Sequence Integration'
method) that is found to be consistent with SED-fitting estimates
after accounting for age uncertainties. The trends shown in the figure
are extrapolations down to the mass range corresponding to the dIrr
for which CMD-based estimates are available. These are shown as black
symbols with error bars in the figure. The right panel shows
SED-fitting estimates for galaxies in different mass bins, compared to
the CMD estimates by \citep{Weisz_etal_2011}. The figure shows that
estimates based on the CMD are largely inconsistent with those
obtained extrapolating results based on SED-fitting techniques. These
results are confirmed by recent full spectrum fits of high
signal-to-noise data in the central regions of nearby disk galaxies
\citep{Sanchez-Blazquez_etal_2011}, as well as by a detailed analysis
of the star formation and chemical enrichment histories based on very
deep data for isolated Local Group galaxies
\citep{Hidalgo_etal_2011}. Therefore, CMD estimates (that are likely
more accurate than those based on SED-fitting techniques) suggest that
dwarf galaxies form large fractions of their stars at $z>1$, as
massive galaxies do. The right panel of Figure~\ref{fig:leitner} shows
that this makes the trends with galaxy stellar mass non-monotonic,
unless there are other systematics in the modelling not yet under
control. In the extreme case that all SED-fitting estimates need to be
corrected, this could even cancel any trend with galaxy stellar mass,
and possibly remove some of the problems discussed above!

Given the strong constraining power of these observables, a better
understanding of age uncertainties in this mass range is clearly
needed.

\section{Question 3 - Does the central-satellite division provide the right 
  framework to study galaxy evolution?}

The classification of galaxies into central and satellites provides an
intuitive framework to study the effects of large scale environment on
the evolution of galaxies.  The framework assumes that all galaxies
form at the center of dark matter halos. Galaxies that are located at
the center of the most massive, dominant halo are defined as
``central'' galaxies, and those that are in bound subhalos of more
massive halos are defined as ``satellite'' galaxies.  In principle,
the framework allows us to determine whether it is the galaxy's own
halo, or the halo of another galaxy that plays the dominant role in
its evolution.  

\begin{figure}
\begin{center}
\includegraphics[width=1.0\textwidth]{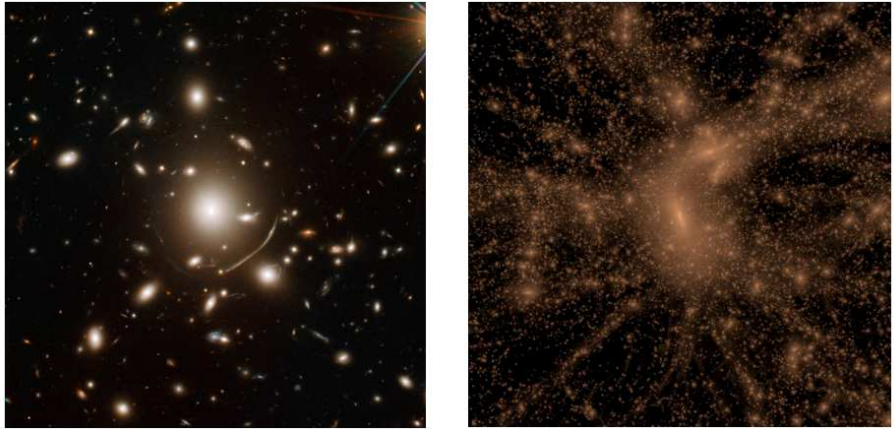}
\caption{Left panel: HST image of Abell 383 at $z =$ 0.1887 (Credit: NASA, ESA,
  J. Richard (CRAL) and J.-P. Kneib (LAM)).  The brightest cluster galaxy is at
  the center of the galaxy distribution and illustrates the basis of the
  central/satellite framework.  Right panel: Dark matter distribution from the
  {\it Via Lactea} simulation \citep{Diemand2007} of a Milky-Way-mass dark
  matter halo (Credit: Diemand, Kuhlen, Madau).  Like the cluster, the
  definition of the central halo and the satellite subhalos seems unambiguous.}
\label{fig:abell}
\end{center}
\end{figure}

In simulations, the distinction between a dominant halo and the
  other bound substructures is clear at all mass scales (the right
  panel of Figure~\ref{fig:abell} shows a snapshot at z=0 of a Milky
  Way-like halo - it is well known that, in a hierarchical universe,
  galaxies are scaled versions of galaxy clusters
  \citep{Moore_etal_1999}). In addition, in all theoretical models of
  galaxy formation and evolution, central galaxies are bound to be
  ``special'': these are the only galaxies onto which gas that is
  shock heated to the virial temperature of dark matter haloes cools
  radiatively. Therefore, it is undoubtedly useful to split observed
  galaxy samples into central and satellite galaxies. But how
  accurately can we correctly identify galaxies as either centrals or
  satellites?

In rich galaxy clusters (see e.g. Figure~\ref{fig:abell}, left
  panel), the brightest galaxies are typically located at the centre
  of a distribution of lower luminosity satellite galaxies. The
  identification of the `brightest cluster galaxy' (BCG) is relatively
  straightforward in most of the cases, but not always: Coma has two
  very bright members located in the proximity of the peak of the
  X-ray emission. 

\begin{figure}
\begin{center}
\includegraphics[width=1.0\textwidth]{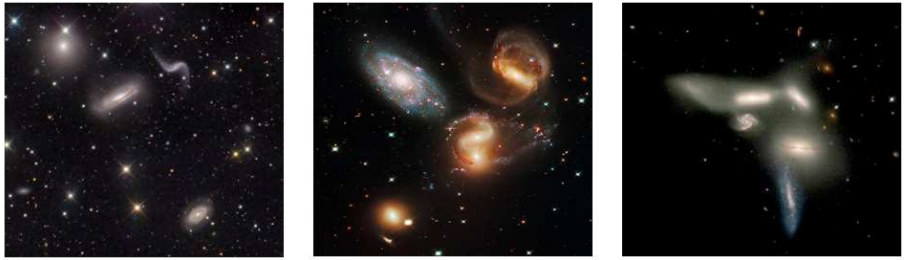}
\caption{From left to right, HST images of Hickson Compact Groups 44,
  92, and 79.  Unlike the examples in Figure~\ref{fig:abell}, where it
  is unambiguous which galaxies are centrals and which are satellites,
  it is not clear which galaxies of these groups should be considered
  the centrals and which should be considered the satellites.  This is
  because there is not a particular galaxy that is clearly the most
  massive and at the center of the overall distribution.  The Hickson
  groups are more spatially compact than most groups so are useful for
  illustration; however, most galaxies do live in groups of similar
  total numbers of galaxies as the Hickson groups.  Therefore, it is
  not clear that the central/satellite framework is appropriate for
  the study of most galaxies.}
\label{fig:hickson}
\end{center}
\end{figure}

 As highlighted by P. van Dokkum at the Lorentz meeting, while the
framework is undoubtedly an intuitive way to approach the question of
large scale environment, and there are clearly scales on which the
framework seems to make sense (e.g. the scale of rich galaxy
clusters), there may also be scales where it may be less functional.

For example, Figure~\ref{fig:hickson} shows images of three Hickson
compact groups \cite{Hickson1982}.  While these are clearly bound
groups, it is not at all clear which galaxy should be designated the
``central'', and which should be the ``satellites''.  It is worth
noting that groups as compact as the Hickson groups are rare, and that
many groups have clear dominant central galaxies such as seen in
galaxy clusters.  However, the existence of the Hickson groups, as
well other groups with similar mass ratios of galaxies, but less
compact distribution, do show that the central/satellite framework
will clearly have limitations in certain cases.

\begin{figure}
\begin{center}
\includegraphics[width=0.9\textwidth]{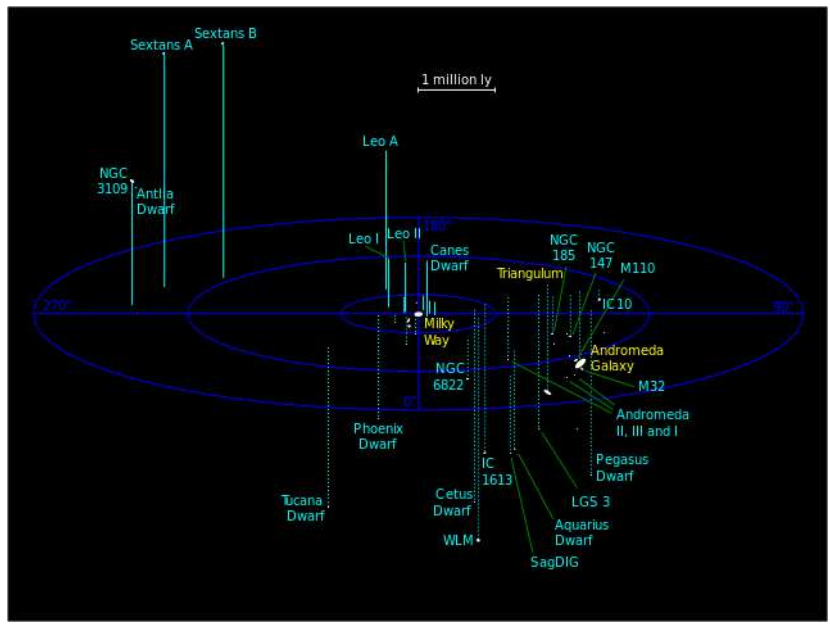}
\caption{Schematic layout of the Local Group (Credit: Richard Powell).  Both
  the Milky Way and M31 are centrals of a system of smaller satellites.
  However, many of the typical central/satellite algorithms when applied to the
  system would likely classify M31 as the central, and the Milky Way (and its
  satellites) as satellites of M31.  It is possible may cause erroneous
  conclusions to be drawn, as galaxies such as the LMC and SMC cannot be
  influenced by their ``central'' in the same way that M32 is. }
\label{fig:lg}
\end{center}
\end{figure}

Along the same lines, Figure~\ref{fig:lg} shows a schematic layout of
galaxies in the Local Group.  While it seems intuitive that the Milky
Way and M31 should both be considered centrals of a large population
of their own satellites, it is likely given the distance and
relatively low peculiar velocity between the two that they would be
considered to belong to the same halo in many satellite/central group
finders.  If so, then which of the Milky Way or M31 should be
considered the central in the Local Group, and which the satellite?
M31 is a more massive galaxy, so likely it would be designated as the
central; however, from Figure~\ref{fig:lg}, it is quite clear that the
relationship of M31 to its nearest satellites must be quite different
than its relationship to the Milky Way.  Likewise for the relationship
of Milky Way and its satellites.  We have to consider that the
relationship of the Milky Way to M31 cannot be the considered on the
same grounds as the relationship of galaxies such as the LMC and SMC
to it.  The situation would be worse if the Milky Way is assigned as a
satellite of M31. Then, all of the Milky Way's satellites would also
be assigned to be satellites of M31.  Herein lies a potentially
significant problem, as it is clear that the evolution of the LMC and
SMC must be more governed by the halo of the Milky Way, not M31, and
this is lost in the central/satellite framework.

Configurations such as the Local Group, with two massive galaxies in
close proximity may not the most common type of group in most group
catalogs; however, the Local Group is a useful illustration of two
potential issues with the central/satellite framework.  The first is
that every bound system can have only one central galaxy, with the
rest of the galaxies being satellites of that central.  The concern
here is that we may be losing our ability to understanding the physics
of galaxy evolution in systems similar to the Local Group if we are
assigning the SMC and LMC satellites of M31, a galaxy that clearly
does not dominate their future evolution.

One of the current challenges is that at present, the data are not
deep enough to be able to resolve the finer structure in satellites,
such as in the Local Group configuration.  For example, in the
well-used \cite{Yang2007} catalog, $\sim$ 85\% of galaxies that live
in groups of at least 2 galaxies, have only 2 galaxies in the
``group''.  This is because the absolute magnitude limit is M$_{r}$ =
-19.5, which is only about 1 magnitude fainter than the characteristic
magnitude at this redshift, M$^{*}_{r}$ = -20.7,
\cite{Monterodorta2009}.  At this depth, the Milky Way and M31 system
would appear as a group of two galaxies using this group finder.
Therefore it is possible that the central/satellite framework may be
failing to capture the interesting effects of environment on the
evolution of many galaxies.  If so, then it clearly demands an answer
to the question of if the central/satellite division provides the
right framework to study galaxy evolution?

It is not an easy question to answer.  One study that attempted to
quantify what information is gained or lost using the framework is by
\cite{Phleps2014}.  They compared the fractions of red satellite
galaxies as a function of both local density estimators, and with the
central/satellite group catalog of \cite{Yang2007}.  They found that
at high densities and high halo masses the results generally
converged; however, at lower densities and lower halo masses they did
not.  They attributed this difference to the challenges of assigning
galaxies as either centrals or satellites when halo masses are low.
\newline\indent Speculating, it would seem that the framework does
make sense when the mass ratio between the central and the satellites
is large, and we resolve much lower mass satellites.  That works with
the current data in systems such as massive galaxy clusters, where the
central is very massive, but the satellites are still massive enough
that many are detected down to the limit of the SDSS.  Where it seems
to be non-intuitive is when the mass ratio of the central and any of
the satellites approach unity, and we cannot detect the fainter
satellite population.  The Local Group is a good example of this.
Things are also problematic when there is no galaxy that is clearly
central spatially in a group, such as the Hickson compact groups.  The
question that needs to be answered in a quantitative way is, what
information are we really loosing by forcing systems like the Local
Group into the central/satellite division?  How can we know what
information is lost?  What is the most important information?

While some attempts have been made, clearly more work is needed to
address these questions.  Whatever the answer, it is likely that the
framework will remain popular for many types of analysis for the
foreseeable future.  The advantage of the satellite/central framework
is that it is a tool that allows simulations and data to be put in a
similar classification scheme, which thereby allows us to make
testable predictions of environmental effects on galaxy evolution
using simulations.  This utility likely outweighs many of the possible
issues, and therefore it is likely to remain for lack of better
alternatives.  Furthermore, deeper surveys such as GAMA are coming
online, and they will permit a more robust definition of
central/satellite than can be done with the current SDSS catalogs.

As pointed out by S. McGee at the meeting, the use of light cones
built from simulations or semi-analytic models allows a direct
comparison between model predictions and observational data to be
carried out. In fact, an increasing number of studies are taking
advantage of ``mock catalogues'' to interpret data in the framework of
theoretical models
\citep{Stringer_etal_2009,delaTorre_etal_2011,Cucciati_etal_2012}. In
principle, these comparisons can circumvent the need to utilize the
central/satellite framework. In practice, however, this division is
still used when interpreting the observational data in the framework
of the models.


\section{Question 4 - We understand which processes affect galaxies in different
  environments.  Do the details matter?}  It has been known for a long
time that the local and large scale environment play an important role
in determining many galaxy properties. The milestone paper in the
subject is probably that of \citep{Dressler_1980}. Theoretical studies
on this topic started early on and indicated the existence of a
plethora of physical processes that can influence the evolution of
galaxies in different environments \citep[][and references
  therein]{DeLucia_2007}. The efficiency and influence of each of
these processes has been studied in detail using dedicated numerical
experiments. In the real Universe, however, physical processes act in
a complex network of actions, back-reactions and self-regulations that
makes their relative importance difficult to quantify. So, while we
can say (citing M. Balogh at the Lorentz meeting) that we know which
processes affect the evolution of satellite galaxies (ram-pressure,
harassment, strangulation, tidal stripping, and assembly bias are all
at work), we certainly do not understand how these processes combine
to establish the detailed trends we observe. Therefore, the obvious
answer to this question is: yes, the details do matter because, in
order to understand how galaxies evolve in different environments, we
need to quantify the relative role of different physical processes.

The completion of large spectroscopic surveys both in the local
Universe and at higher redshift has allowed a detailed quantification
of the physical properties of satellite galaxies in a
multi-dimensional space (stellar mass, halo mass, redshift, radial
distance from the cluster centre). As mentioned in the previous
section, many key results concerning the evolution of galaxies as a
function of the environment have come from the application of the
central/satellite framework to SDSS data. These studies have
highlighted that satellite galaxies are on average redder and less
frequently star-forming than their centrals, regardless of the central
halo mass and the satellite stellar mass \citep[e.g.,][]{Weinmann2006,
  vandenbosch2008,Weinmann2009}. The central/satellite framework also
revealed the puzzling issue of the ``galaxy conformity''
\citep{Weinmann2006}, which is that satellites tend to somehow know
about the star formation properties of their centrals.  Star-forming
centrals have a significantly higher fraction of late-type satellites
than haloes with an early-type central galaxy.  As of yet, the reason
for the galaxy conformity is not well understood.  It has been
speculated that it may arise from an assembly bias, with quiescent
centrals having accreted their satellites earlier, hence they have
been affected by environmental processes longer
\citep{Yang2006}.  It has also been suggested that quiescent
centrals inhabit more massive halos than star-forming centrals of
similar stellar mass \cite{Wang2012}.  It is also possible that
some of the conformity may occur because of the misclassification of
some satellites as centrals \citep{Kauffmann2013}.

\begin{figure}
\begin{center}
\includegraphics[width=0.9\textwidth]{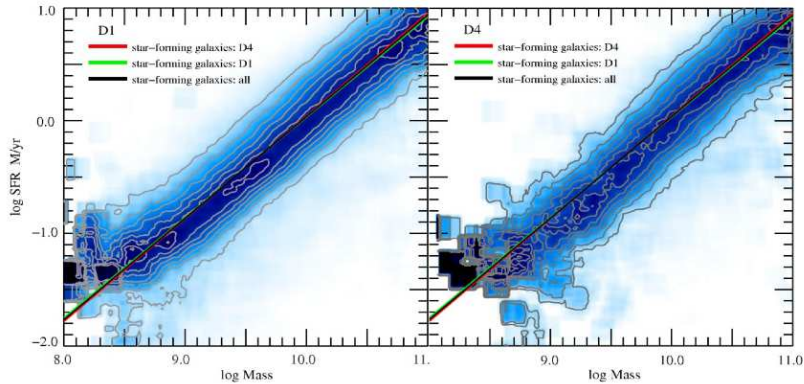}
\caption{Figure from \citep{Peng_etal_2010}. The panels show the
  relation between star formation rate and stellar mass for
  star-forming SDSS galaxies in the low density D1 quartile (left) and
  high density D4 quartile (right). The three almost indistinguishable
  lines, reproduced on both panels, show the fitted relation for all
  galaxies, and for those in the D1 and D4 density quartiles.}
\label{fig:peng}
\end{center}
\end{figure}

One recent results on the evolution of satellite galaxies that was heavily
discussed at the Lorentz meeting is shown in Figure~\ref{fig:peng} from
\citep{Peng_etal_2010}. The figure shows how the star formation rates of star
forming galaxies from the SDSS depend on galaxy stellar mass, in low (left
panel) and high (right panel) density regions.  Assuming that high density
  regions mainly contain satellite galaxies, the figure shows that active
satellite galaxies form stars at a rate that is very similar to that of central
galaxies of the same mass, suggesting that the transition from active to
passive must be very rapid. Considering group catalogues based on SDSS Data
Release 7, \citep{Wetzel_etal_2012} have studied the specific star formation
rate distribution of satellite galaxies and its dependence on various physical
properties. All galaxies (both centrals and satellites) exhibit a similar
bimodal distribution of specific star formation rates. The distribution, and in
particular its minimum (corresponding to the so-called `green valley'), appear
not to depend strongly on stellar mass, halo mass, and distance from the halo
centre. This puts important constraints on the efficiency and time-scales of
the processes affecting the evolution of satellite galaxies. In particular, (1)
satellite galaxies must have evolved as central galaxies (forming stars at very
similar rate) for several Gyrs; (2) once `quenching' of star formation begins,
this must occur on a rapid timescale in order to avoid an excess of galaxies at
intermediate values of sSFR; (3) there is no minimum halo mass for
satellite-specific processes. Several other studies have argued for relatively
long time-scale for the suppression of the star formation rates in satellite
galaxies \citep[for example][and references
  therein]{Balogh_etal_2000,McGee_etal_2011,DeLucia_etal_2012}. This turns out
to be a very tough constraint for theoretical models.

As discussed above, the basic working hypothesis of modern theories of
galaxy formation is that, when a galaxy is accreted onto a larger
halo, the cold gas supply can no longer be replenished by cooling,
because of the stripping of the hot gas reservoir
\citep{Larson_etal_1980}. This leads to the exhaustion of star
formation in the galaxy, on a time-scale that depends on how fast the
stripping of the hot gas associated with the infalling galaxy is, how
efficient the stellar feedback is, and what is the fate of the reheated
and ejected gas. For all of these processes, our understanding is
incomplete (this is especially true for what concerns the stellar
feedback), and current implementations all induce a very rapid decline
of the star formation histories of satellite galaxies, contributing to
create the excess of red and passive galaxies discussed above (see
Question 2). A more gradual stripping, that follows the stripping of
the dark matter substructures or simple models inspired by numerical
simulations, does not provide yet a satisfactory agreement with
observational data. It remains unclear how satellite galaxies can
sustain significant levels of star formation for several Gyrs, and if
this can be achieved by simply relaxing the assumption of
instantaneous stripping of the hot gas reservoir associated with
galaxies when they are accreted. In addition, it should be considered
that at least part of the problems with satellite evolution might be
related to the fact that current models could predict the wrong
properties for central galaxies at the time of accretion.

In general, a detailed quantification of the influence of different
environmental physical processes is complicated by the fact that one
cannot easily separate ``nature'' from ``nurture''. According
to the current paradigm for structure formation, dark matter collapses
into haloes in a bottom-up fashion: small systems form first and
subsequently merge to form progressively larger systems. As structure
grows, galaxies join more and more massive systems, therefore
experiencing a variety of environments during their lifetime. In this
context, the nature-nurture debate appears to be ill posed, as these
two elements of galaxy evolution are inevitably and heavily
intertwined \citep{DeLucia_etal_2012}.

\section{Ways forward}

In the previous sections we have highlighted four open questions
related to the broad subject of galaxy evolution. What can be done in
order to make progress on these questions? Below we discuss what we
believe are the main steps to be taken in order to make progress on the
questions discussed above.

\subsection{Future Improvements in Data for Stellar Mass Determination}

One of the fundamental parameters needed to measure the stellar mass
of a galaxy is the redshift.  For distant galaxies, where
spectroscopic redshifts are currently few, usually the broadband SED
is used simultaneously for redshift determination and for fitting of
the stellar mass.  Clearly if there are large systematic errors in
determining the photometric redshifts (either from the method, or
template set), then the stellar masses will also be incorrect.  We
know this happens at some level for all samples, as even the very best
sets of photometric redshifts determined at high redshift
\citep[e.g.,][]{Ilbert2009,Whitaker2011,Ilbert2013,Muzzin2013a}
typically have catastrophic outlier fractions (when compared against
even very limited sets of spectroscopic redshifts) of 1\% -- 5\%.
Therefore it is clear that obtaining spectroscopic redshifts for large
samples of distant galaxies is a very straightforward and tractable
way of improving our current measurements of stellar masses.

Large optical spectroscopic redshift surveys of the distant Universe
have been done (e.g., DEEP2, zCOSMOS, GOODS, VVDS, PRIMUS), and ever
new ones are coming online (e.g., VIPERS, VUDS).  These have been
extremely useful for verifying photometric redshifts (and thereby
stellar masses).  They have, however, all selected targets based on
optical selection criteria.  As far as the authors are aware, a
spectroscopic redshift survey which aims to be complete in {\it
  stellar mass} (this would require a selection in the rest-frame
  optical) up to a given redshift does not yet exist, and could be
very useful for a determination of the stellar mass function that is
robust to photometric redshift uncertainties.

The 8m class telescopes are now commissioning, or have already
commissioned their NIR multi-object spectrographs (e.g., FMOS,
MOSFIRE, KMOS, MMIRS, FLAMINGOS2).  Spectroscopic redshift surveys
from these will no doubt help us push to higher redshift, and allow us
to more easily access spectroscopic redshifts for the most
intrinsically red galaxies.

Spectra will also be invaluable for better determining stellar masses
via fitting of the spectra.  As shown by
\citep[e.g.,][]{Kauffmann2004} with SDSS data, more information about
the star formation history can be determined from spectra than broad
band photometry alone, and much of it can be done in a way that is
independent of dust, a significant complicating factor in the
determination of mass-to-light ratios.  Rest-frame optical spectra
with high signal-to-noise (S/N) can also be used to determine stellar
metallicities \citep[e.g.,][]{Thomas2005,Gallazzi2005,Thomas2010},
thus removing an additional assumption that must be made in the case
of broadband photometry.  In the case of very high S/N spectroscopy,
dynamical masses can be determined from the widths of the absorption
lines.  While these suffer from the fact that they must be corrected
to a fixed physical size, and measure the total enclosed mass (dark,
stellar, and gas), they do provide extremely valuable upper limits of
what the total mass can be.  

Another fundamental limitation in our modelling of stellar masses is
the correct dust attenuation law to use.  Only a few have been
measured directly, such as the LMC, SMC, Milky Way, and the Calzetti
law for starburst galaxies.  These are notably different from each
other, and given they are determined locally, it is unclear whether
they apply in the distant Universe where star forming regions are
likely quite different, and metallicity is lower.  Efforts have been
made to empirically determine new dust laws from galaxies
\citep[e.g.,][]{Kriek2013}, quasars \citep[e.g.,][]{Maiolino2001,
  Gallerani2010} and from the power-law SEDs of gamma ray bursts
\citep[e.g.,][]{Kruhler2011, Schady2012}.  Additional empirical
determinations of the dust attenuation law for distant galaxies of
various classes and stellar masses would be extremely valuable for
determining better masses, and is also potentially tractable in the
future.

Lastly, one approach that has only begun to be exploited is to model
the stellar masses of galaxies in a spatially resolved way.  This
could be extremely valuable for removing both the universal ``dust
screen'' assumption, where the dust obscures the entire stellar
population (young and old) in the same manner, and for removing the
need for the entire galaxy to be fit to a single parametric SFH.
Efforts to do this both locally
\citep[e.g.,][]{Welikala2008,Zibetti2009} and in the distant Universe
\citep[e.g.,][]{Elmegreen2009,Forsterschreiber2011,Wuyts2012} have
shown that this is only modestly different than standard assumptions
for most galaxies, but does substantially change the stellar masses
for small subsets of the population.  Overall, this would be a
positive step forward, and the upcoming CALIFA \citep{Sanchez2012} and
MANGA\footnote{See http://www.sdss3.org/future/manga.php} surveys will
provide high-quality data for this type of modelling for hundreds to
thousands of galaxies.

\subsection{Future Improvements in Models for Stellar Mass Determination}
\label{sec:futurespmodels}

It is well documented that most of the well-used SPS models
\citep[e.g.,][]{Bruzual2003,Maraston2005,Conroy2009} produce different
stellar masses with the same observational data.  This is clearly
problematic and underlies the uncertainties in the process of SPS.
Here we discuss some straightforward improvements that could be made.
Much of this discussion has been influenced by the review paper by
\cite{Conroy2013}, and we refer the reader to that paper for a more
in-depth discussion of the topic.

Currently, one of the most heavily debated issues in the SPS community
is the treatment of the TP-AGB phase of stellar evolution.
It was pointed out by \cite{Maraston2005} that at that time, models may not treat this short-lived but bright phase of stellar evolution correctly, and \cite{Maraston2005} provided a new set of models with an updated treatment of TP-AGB evolution. Later studies pointed out that this
treatment might not be ideal as it over-predicts the amount of
rest-frame NIR flux for post-starburst galaxies
\citep{Kriek2010,Zibetti2013}. \cite{Conroy2009,Conroy2010b} include
the strength of the TP-AGB as a variable as part of their SPS models,
and \cite{Bruzual2007} continue to update their models to reflect the
best-possible treatment.  It is now clear that convergence on the
treatment of this phase of stellar evolution is a mandatory ingredient
for reducing the systematic uncertainties in deriving stellar masses.

Several other obvious aspects of stellar evolution are also missing
from current codes, and their implementation would also be
useful. These include rotation of the stars (that has the effect of
lengthening the main-sequence lifetime of stars by $\sim$ 25\%
\citep{Maeder2000}), and binary star evolution. As suggested by
\cite{Conroy2013}, one of the most obvious elements missing from the
most-used SPS models (PEGASE excepted) is the inclusion of emission
lines in the synthetic spectra.  Emission lines contribute to the
broad band flux of galaxies, and because of the short timescales of
star formation they probe (compared to the UV), they give valuable
information on the SFH.  Contamination of broad band fluxes from
emission lines is not a significant issue in the local Universe, where
the typical equivalent widths (EWs) of galaxies are of order of a few
to tens.  However, recent work has shown that strong emission lines,
with EWs up to $\sim$ 1000\AA~can be seen for very young galaxies at
$z > 4$ \citep[e.g.,][]{Schaerer2009,Stark2013,Labbe2013,Smit2014},
and this can lead to a significant overestimation of their stellar
masses.  It has also been shown by \cite{vanderwel2011,Maseda2013}
that dwarf galaxies at $z \sim$ 2 have extreme EWs, up to $\sim$
1000\AA. 

Finally, dust remains a major issue.  One clear way forward in this
regard is to produce SPS codes that simultaneously take account of the
emission from stars and its subsequent absorption and re-emission at
longer wavelengths by dust.  If FIR data is available, then dust
absorption and emission can be accounted for in a self-consistent way.
In particular, this could be very helpful in improving on our
understanding of dust geometry and attenuation law on a
galaxy-by-galaxy basis.  Some such codes have been developed such as
MAGPHYS \cite{dacunha2008} and CIGALE \cite{Noll2009} and their
refinement is clearly an avenue forward in terms of improving our
ability to model stellar masses.

The last consideration that has not been discussed significantly in
this article (neither at the meeting) is the IMF, that can vary
\citep[e.g.][]{Gunawardhana_etal_2011,Cappellari_etal_2012,Conroy_etal_2013}. Given
the challenges of determining the IMF for any galaxies other than
locally, the IMF may prove to be a fundamental limitation in our
ability to measure stellar masses.

\subsection{Future improvements in Theoretical Models of Galaxy Evolution}

The last decade has witnessed enormous progress in our ability to
model the formation and the evolution of galaxies in a cosmological
context. The advent of more and more powerful computers, and the
development of more sophisticated algorithms has allowed us to carry
out simulations with increasing resolution, and including sophisticated
modelling of various physical processes. Dark matter substructures
are nowadays routinely identified and their history robustly
tracked. These ``merger trees'' can be used to construct halo occupation
and abundance matching models, as well as a basis for semi-analytic
models of galaxy formation. Parallel effort with hydrodynamical
simulations has allowed us to understand, at least qualitatively, the
role of different physical processes and their importance at different
scales. 

Despite progress, however, some persistent problems have been
identified. In particular, all models over-predict the number
densities of low to intermediate stellar mass galaxies and
under-predict the fraction of active satellite galaxies. As discussed
above, these two failures might well be different manifestations of
the same problem. The current wisdom is that the solution of the
problem lies in a physical process that is able to break the
parallelism between mass growth and halo growth. It remains to be seen
if this can be achieved by modifying the stellar feedback
scheme. Likely, it is the modelling of the ``self-regulation'' between
the star formation and the stellar feedback that should be improved,
in particular for low-mass galaxies. In fact, dwarf galaxies in the
Local Group appear to have had ``bursty'' and stochastic star
formation histories \citep{Grebel_2000} - a constraint that is very
difficult to reproduce in the framework of available simulations and
galaxy formation models. In this framework, gas recycling and the
differential ejection of metals and gas are important processes to
understand. Both detailed numerical simulations and dedicated
observations are required to this end. Parallel effort in developing
self-consistent phenomenological models of galaxy evolution
\citep[e.g.][]{Wetzel2013,Berhoozi_etal_2013}, as well as in testing
non-standard dark matter models and/or cosmologies
\citep{Menci_etal_2012,Lovell_etal_2012,Fontanot_etal_2013} will be
important to understand what is required as an addition to our current
theory of galaxy formation and evolution.

From the numerical point of view, published work on the role of
environment on galaxy evolution is still largely focused on massive
galaxy clusters (simulations of ram-pressure stripping in poor groups
have been carried out by \citep{Marcolini_etal_2003} and by
\citep{Roediger_Hensler_2005}; the effects of both tides and
ram-pressure have been analysed for the specific case of the Local
Group by \citep{Mayer_etal_2006}). Yet, as discussed above, groups
likely represent the most common environment that galaxies
experience\footnote{How much time galaxies spend in a ``group''
  environment will depend on the galaxy stellar mass and on the actual
  definition of galaxy group \citep[see e.g.][]{DeLucia_etal_2012}.}.
The situation is slowly improving \citep[see
  e.g.][]{Villalobos_etal_2012}, but additional work on detailed
controlled numerical experiments (including the evolution of the
gaseous components) is needed in order to quantify the importance of
different physical processes, at the typical velocity dispersions of
galaxy groups. This will also help us understanding to what degree the
problem with satellite galaxies is related to a poor treatment of
environmental processes.

In order to keep up with the imminent observational revolution that
will happen in the next future, models need to be further extended:
MANGA is scheduled to begin in the Fall of 2014, and it will
ultimately allow resolved spectral measurements for about 10,000
galaxies in the nearby Universe. SKA (and its precursors) will measure
the abundance of HI in the Universe and its evolution, while ALMA will
soon begin to probe the molecular gas contents in high-redshift
galaxies. In order to make the best use possible of the overwhelming
amount of data that will come in the next future, our theoretical
tools need to be updated including an explicit treatment of the
transition from atomic to molecular gas, and (possibly) an explicit
modelling of spatially resolved physical properties. Work has started
in this direction
\citep{Fu_etal_2010,Lagos_etal_2011,Lagos_etal_2012}.

Finally, as pointed out in the final discussion session at the Lorentz
meeting by C. Conroy, metals (including detailed abundance ratios)
still represent a underutilized, yet very powerful constraint to
models of galaxy formation and evolution \citep[see
  e.g.][]{Pasquali_etal_2010,DeLucia_Borgani_2012}. Therefore, the new
generations of galaxy formation models and simulations should all
include an explicit and detailed treatment of the chemical evolution
of different species.

\vspace{1truecm}

\noindent {\bf Acknowledgements} We acknowledge the Lorentz Centre for
giving us the opportunity to organize a successful and very
stimulating workshop, and for making the organization process so
easy. We are grateful to Marijn Franx for financial support to
our workshop, and to the editors of New Astronomy Reviews for accepting
our contribution. GDL acknowledges financial support from the European
Research Council under the European Community's Seventh Framework
Programme (FP7/2007-2013)/ERC grant agreement n. 202781. \\
\noindent Last but not least, we thank all the participants for making
such a great workshop!

\section*{References}

\bibliography{mybibfile}

\end{document}